\documentclass[10pt]{article}
\usepackage{geometry}               
\usepackage[parfill]{parskip}
\usepackage{amssymb}

\usepackage{amsmath}
\newcommand{\sech}{\mbox{sech} }

\title{\bf Recurrence relations and Path Representations of Matrix Elements of an Algebra related to su(2) and su(1,1)}

\author{C. V. Sukumar \\{\em Wadham College,}\\{\em University of Oxford, Oxford OX1 3PN, U.K. }
}

\begin{document}
\maketitle

\begin{abstract}
An algebra which may be used to provide a unified description of the simple harmonic oscillator and the angular momentum algebras and a class of other algebras defined on a semi-infinite state space is identified. A normal ordered representation of a Unitary operator $U$ constructed from the generators of the algebra, which is a generalization of the Baker- Campbell - Hausdorff relation for Lie algebras, is given. It is shown that the normal ordered representation of $U$ may be used to calculate expectation values which are functions of the parameters used to construct the operator. The functions so constructed satisfy certain recurrence relations and the entire set of functions may be interpreted in terms of diagrams similar to the Pascal triangle for binomial coefficients. Coherent states, squeezed states and rotation matrices of the angular momentum algebra emerge as special cases. 
\end{abstract}

\section{Introduction}
We consider a set of operators defined on a set of orthogonal basis vectors of a separable, finite or infinite dimensional complex vector space $H$ with vectors $|j\rangle$, where $j=0$ refers to the vacuum vector, positive integer values of $j$ refer to 'particle' vectors and negative integer values of $j$ refer to 'hole' vectors. Let the hermitian adjoint of an operator $A$ in the vector space be defined by $<m|A|n> = <n|A^{\dagger}|m>^{\star}$. Let the matrix representations of the operators $L$, its hermitian adjoint $L^{\dagger}=R$  and their commutator $S=[L,R]$ in the vector space $H$  be defined by
\begin{align}
L_{jk}\ &= \ \lambda_{j}\ \delta_{j,k-1}\ \ ,\ \ R_{jk}\ =\ \lambda_k\ \delta_{j,k+1}\ ,\ \ j,k=0,\pm1,\pm2,.... \\
S_{jk}\ &= \ \ (\lambda^2_j\ -\lambda^{2}_{j-1})\ \delta_{jk}\ \label{}
\end{align}
where $\lambda_{j}$ are real numbers. The operators $R,L$ and $S$ are the raising, lowering  and number operators of the algebra. Since 
\begin{equation}
[L,S]_{jk}\ =\ \lambda_j\ (\lambda^2_{j+1}\ +\ \lambda^2_{j-1}\ -\ 2 \lambda^2_j)\ \delta_{j,k-1} \label{}
\end{equation}
the algebra closes if the parameters $\lambda_j$ have $j$ dependence of the form
\begin{equation}
\lambda^2_j\ =\ \sigma\ (\alpha+j)\ (\beta+j)     \label{}
\end{equation}
so that
\begin{equation}
\lambda^2_{j}\ -\ \lambda^2_{j-1}\ =\ \sigma (2j\ -1\ +\alpha\ +\beta),\ \ \quad  \lambda^2_{j+1}\ +\ \lambda^2_{j-1}\ -\ 2\ \lambda^2_j\ =\ 2\ \sigma  \label{}
\end{equation}
 where $\alpha$, $\beta$ and $\sigma$ are real numbers. The resulting closed algebra 
\begin{equation}
 [L,R]\ =\ S\ \ ,\ \ [L,S]\ =\ 2\ \sigma\ L\ ,\ \ [S,R]\ =\ 2\ \sigma\ R \ \label{eq:Al}
\end{equation}
can be identified with one of the well known algebras depending on the choice of $\sigma$. If $\sigma$ is positive the algebra may be identified as an su(2) algebra or its complex extension $A_1$. If $\sigma$ is negative the algebra may be identified as an su(1,1) algebra.

Of the three parameters present in the algebra $\sigma$ is a scale parameter while $\alpha$ and $\beta$ provide different possibilities for the realization of the algebra depending on whether $(\alpha,\beta)$ are positive or negative, integer or non-integer. It is possible to realize the algebra in finite or infinite or semi-infinite dimensional vector spaces by suitable choices of the parameters.  Three familiar realizations of the algebra in (\ref{eq:Al}) can be identified. (We set $\hbar =1$ throughout this paper).

${\bullet}$  The creation and annihilation operators $a^{\dagger}$ and $a$ of the Simple Harmonic Oscillator ( hereafter referred to as SHO)  obeying the commutation relation $[a,a^{\dagger}]=1$ when $2L=a^2$ and $2S=(a^{\dagger}a+aa^{\dagger})$ is an example of the algebra with $\sigma=+1$ defined on semi-infinite dimensional vector space. This possibility corresponds to choosing either $\alpha=2\beta=1$ or $2\alpha=\beta=1$ and either all the even number particle states or all the odd number particle states of the SHO provide an example of a realization of the algebra.

${\bullet}$  If $L =a,\ R = a^{\dagger},\ \alpha=1, \beta\rightarrow \infty$, $\sigma \rightarrow 0$ and $\sigma\beta\ \rightarrow 1$ the resulting algebra is again the algebra of the SHO and all the number states of the oscillator taken together provide another semi-infinite dimensional realization of the algebra. 

${\bullet}$  A finite dimensional example is the angular momentum algebra whose generators ${\sqrt{2}} J_{\pm} = J_x\pm i J_y$ and $J_z$ satisfy the commutation relation $[J_x,\ J_y]\ =\ iJ_z$ and its cyclic permutations and can be mapped on to the $RLS$ scheme by $R=J_+,\ L=J_-,\ S= -J_z$ with $\sigma=-\frac{1}{2},\ \alpha=n,\ \beta=-n-1$ and different choices of integers $n$ provide different finite dimensional realization of the algebra corresponding to different values of $J^2=n(n+1)$. Half-integral angular momenta can also be accommodated in this scheme by switching to the basis vectors $|N\rangle$ with $N$ taking only half-integral values instead of integral values and finite dimensional realizations are possible now whenever $\alpha=N$ and $\beta=-N-1$. 

In this study we consider general cases of the realization of the algebra in (\ref{eq:Al}) when $\alpha,\beta$ and $\sigma$ take arbitrary values. When $\alpha$ and $\beta$ take values which are different from integers or half-integers infinite dimensional algebras arise and these algebras may prove useful in new contexts. The su(1,1) and su(2) algebras has played an important role in Mathematical Physics. Various representations of su(1,1) and su(2) have been extensively studied (Barut and Fronsdal 1965, Holman and Biedenharn 1966). The su(1,1) group and its Schwinger representation (Schwinger 1965) have been used in Quantum Optics in connection with the study of parametric amplifiers (Louisell 1977), interferometers and squeezed states (Yurke {\it et al} 1986). The su(1,1) group has been used to provide a unified description bosons, fermions and anyons in the plane and their relation to supersymmetry (Horvathy {\it et al.} 2010) and in the study of the supersymmetry of Poschl-Teller potentials (Correa {\it et al} 2009).

\section{Normal Ordered Representation}

It is well known in Quantum Optics that the squeezed states and coherent states of light may be generated from the vacuum state of a simple harmonic oscillator mode by the action of a Unitary operator acting on the vacuum state. The generators $(R,L,S)$ may be used to construct a Unitary operator $U_1\ =\ \exp(iy(R+L))$ which is a generalization of the Unitary operator arising in the study of the generation of squeezed states and coherent states of light from the vacuum. The normal ordered representation, in which the lowering operators are aligned on the right and the raising operators are aligned on the left and the anti-normal ordered representation, in which the raising operators are on the right and the lowering operators are on the left, of $U_1$ may be found using the method discussed by Truax (1985) and Sukumar (1989) and given in the form 
\begin{align}
U_1\ &=\ \big(\exp(iy(R+L))\big)\ =\ (\exp(iyfR))\ \Big(\sum_j |j\rangle \ (g^{p_{j}})\ \langle j|\Big)\ (\exp (iyfL)) \label{eq:n1} \\
&=\ (\exp(iyfL))\ \Big(\sum_j (|j\rangle\ (g^{-p_{j}})\ \langle j|\Big)\ (\exp(iyfR)) \label{eq:n2} \\
&f\ = \frac{\tanh y{\sqrt{\sigma}}}{y{\sqrt{\sigma}}}\ \ ,\ \ g= \sech\ y{\sqrt{\sigma}}\ \ ,\ \   p_{j}\ =\ (2j-1+\alpha +\beta)  .\label{eq:n3}
\end{align}
This result is a generalization of earlier results which were derived for special values of the parameters ${\alpha, \beta, \sigma}$ and is generalization of the Baker- Campbell Hausdorff relation. The set of functions $G_n$  defined by 
\begin{equation}
G_n\ =\ (-i)^n\  \langle n|\ U_1\ | 0\rangle\ ,\ \ n=0,1,2,...\label{eq:f3}
\end{equation}
may be evaluated using the anti-normal ordered representation of $U_1$ to get
\begin{equation}
G_n(\alpha,\beta;\sigma;y)\ =\ (-i)^n\ \sum_{j=n}^{\infty}\ \frac{(ify)^{(j-n)}}{(j-n)!}\ \langle n| L^{(j-n)} |j\rangle\ g^{(1-2j-\alpha-\beta)}\ \langle j|R^j|0\rangle\ \frac{(ify)^j}{j!}\ .\label{eq:f2} 
\end{equation}
The function in (\ref{eq:f3}) has been considered before ( up to the scaling by ${\sqrt{\sigma}}$ by Ui (1970)).
The matrix representations of the generators $(R,L)$ and the power series expansion of the hypergeometric functions of the second kind (Gradshteyn and Ryzhik 1965) may be used to show that 
\begin{align}
G_n(\alpha,\beta;\sigma;y) &= A_n \Big(\frac{\tanh y{\sqrt{\sigma}}}{{\sqrt{\sigma}}}\Big)^n  (\sech\ y{\sqrt{\sigma}})^{\alpha +\beta - 1}\ _2F_1(1-\alpha, 1-\beta; 1+n ;-\sinh^2 y{\sqrt{\sigma}}) \label{eq:f1}\\
A_0 = 1,\ \ A_n &=\ \sqrt{\sigma^n\ \frac{\Gamma(n+\alpha)}{\Gamma(\alpha) \Gamma(n+1)}\ \frac{\Gamma(n+\beta)}{\Gamma(\beta) \Gamma(n+1)}}\ =\ \frac{1}{n!}\ \prod^{n}_{j=1}\ \lambda_{j-1}\ \ ,\ n=1,2,... \label{eq:f0}
\end{align}
Using $L|n+1\rangle=\lambda_n|n\rangle$ and $R|n\rangle=\lambda_n|n+1\rangle$ and (\ref{eq:f3}) it can be established that $G_n$ satisfy the recursion relation
\begin{equation}
\frac{\partial}{\partial y}G_{n+1}\ =\ \big(\lambda_n\ G_n\ -\ \lambda_{n+1}\ G_{n+2}\big)\ .\label{eq:r1}
\end{equation}
This recursion relation may also be proved directly using the properties of the hypergeometric functions. It is evident that the $G_n$ are symmetric in $\alpha$ and $\beta$. It is clear from (\ref{eq:f1}) that a variety of solutions are possible depending upon the values of $\alpha,\ \beta$ and $\sigma$. We note that the functions $G_n(y)$ are the expansion coefficients that describe a superposition of states arising from the action of the Unitary operator $U_1$ acting on the vacuum state and they depend upon the generators of the algebra and the parameter $y$.

\subsection{Diagrammatic Representations}

The power series representation of the set of solutions $G_n$ may be displayed using a set of nodes on a grid set in the form of a rhombus. Such diagrams in general correspond to generating a 2-dimensional array of points $(j,k)$ such that the sites for which $j+k$ is an even integer are occupied. Weight factors $A_{jk}$ can be associated with the nodes such that $A_{jk}=0$ if $j+k=2m+1$. Using such an array it is possible to study a class of functions of two variables:
\begin{equation}
P(y,q)\ =\ \sum_{j=0}^{\infty}\ \sum_{k=0}^{j}\ A_{jk}\ y^j\ q^k\ =\ \sum_{k=0}^{\infty}\ \sum_{j=k}^{\infty}\ A_{jk}\ y^j\ q^k \ =\ \sum_{j=0}^{\infty}\ y^j\ {\tilde f_{j}}(q)\ =\ \sum_{k=0}^{\infty}\ f_k(y)\ q^k \label{eq:d1}
\end{equation}
where the sum over the functions ${\tilde f_j}$ corresponds to constructing functions based on the weights associated with the nodes in a row and then summing over such functions with weights $y^j$ while the sum over the functions $f_j$ corresponds to constructing functions based on the weights associated with the nodes in a column and summing over such functions with weights $q^k$.  The nodes in the diagrams then code for a particular class of functions of two variables such that the rows define a sequence of polynomials and the columns define a sequence of functions with infinite number of terms in their power series. Diagrams of this kind are relevant for the algebra under discussion for the following reason. The matrix representations of $(R,L)$ considered in this paper have non-vanishing elements only on either the diagonal above or the diagonal below the principal diagonal. Consequently the $k^{th}$ powers of these matrices have non-vanishing elements only along the $k^{th}$ diagonal above or below the principal diagonal. $U_1$ contains all powers of the matrix $M=R+L$ and the matrix elements of powers of $M$ naturally lend themselves to representation in terms of diamond like formations. When the 'hole' states decouple from the 'particle' states one half of the diamond formation is adequate and leads to triangular diagrams. We examine this case first. We consider the case when one of $(\alpha,\beta)$ equals $1$ and the other takes the positive value $p$ to identify some of the features of the scheme introduced in this section.

\subsection{Study of $(\alpha,\beta) =$ (1,\bf p)\ ,\ $\sigma\ =\ 1$}

For this case $\lambda^2_j= (1+j)(p+j)$ which means that $\lambda_{-1}=0$ and as a consequence the matrix representations of $(R,L)$ exhibit a complete decoupling of the 'particle' and 'hole' states and it would be adequate to restrict attention to the set of states $|n\rangle$ with $n$ in the range $[0,\infty]$. The hypergeometric function in (\ref{eq:f1}) simplifies to just 1. In this case a power series representation of the solutions (\ref{eq:f1}) in the form
\begin{equation}
G_n(y) \ = \sum_{m=0}^{\infty}\ (-)^m \ C_{n,m}\ \frac{y^{2m+n}}{(2m+n)!} \label{}
\end{equation}
leads to the recursion relation
\begin{equation}
C_{n+1,m}\ =\ \lambda_n\ C_{n,m}\ + \ \lambda_{n+1}\ C_{n+2,m-1} \label{}
\end{equation}
which may be interpreted in terms of a diagram of the form 
$$
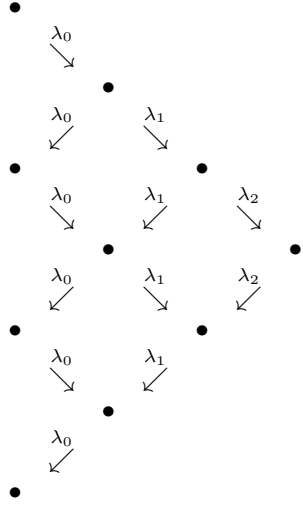
\begin{array}{ccccccccccc}
{\bullet}&&&&&&{  }\\
& \stackrel{\lambda_0}{\searrow}&&&&&{  }\\
&&{\bullet}&&&&{  }\\
& \stackrel{\lambda_0}{\swarrow}&&\stackrel{\lambda_1}{\searrow}&&&{ }\\
{\bullet} &&&& {\bullet}&&{ }\\
& \stackrel{\lambda_0}{\searrow}&&\stackrel{\lambda_1}{\swarrow}&&\stackrel{\lambda_2}{\searrow}&{ }\\
&&{\bullet}&&&&{\bullet}\\
& \stackrel{\lambda_0}{\swarrow}&&\stackrel{\lambda_1}{\searrow}&&\stackrel{\lambda_2}{\swarrow}&{ }\\
{\bullet}&&&&{\bullet}&&{ }\\
& \stackrel{\lambda_0}{\searrow}&&\stackrel{\lambda_1}{\swarrow}&&&{ }\\
&&{\bullet}&&&&{ }\\
&\stackrel{\lambda_0}{\swarrow}&&&&&{ }\\
{\bullet}&&&&&&{ }\\
\end{array}
$$
where the nodes in the columns represent $C_{n,m}$ for fixed $n$ and varying $m$. The values attached to the nodes correspond to the sums arising from the addition of all contributions from paths pointing towards the node, with each path contributing an amount equal to the product of the value of the node at the start of the arrow and the weight factor attached to the arrow. The power series expansion of 
\begin{equation}
G_n(1,p;1;y)\ =\ G_n(p,1;1;y)\ =\ {\sqrt{\frac{\Gamma(n+p)}{\Gamma(n+1)\Gamma(p)}}}\ {\sech^p{y}}\ \tanh^n{y}\ ,\ n=0,1,2,... \label{}
\end{equation}
is related to the values at the nodes of the columns of the diagram given above. The presence of square root factors  in these functions may be removed by defining
\begin{align}
{\tilde G}_n(p;y)\ &=\ {\sqrt \frac{\Gamma(p)\Gamma(n+1)}{\Gamma(n+p)}}\ G_n(1,p;1;y)\ =\ \sech^p\ y\  \tanh^n y  \\
{\bar G}_n(p;y)\ &=\ {\sqrt{\frac{\Gamma(n+p)}{\Gamma(p)\Gamma(n+1)}}}\ G_n(1,p;1;y)\ =\ \frac{\Gamma(n+p)}{\Gamma(p)\Gamma(n+1)}\ \sech^p\ y\ \tanh^n y \label{}
\end{align}
which satisfy
\begin{align}
\frac{\partial }{\partial y}\ {{\tilde G}_{n+1}(p;y)}\  &=\ (n+1)\ {\tilde G}_{n}(p;y)\ -\ (n+1+p)\ {\tilde G}_{n+2}(p;y)\\
\frac{\partial }{\partial y}\ {{\bar G}_{n+1}(p;y)}\  &=\ (n+p)\ {\bar G}_{n}(p;y)\ -\ (n+2)\ {\bar G}_{n+2}(p;y) \label{}
\end{align}
We note that the case ${\bf p=1/2}$ leads to Squeezed states in Quantum Optics and in particular $n=0$ corresponds to the squeezed vacuum in Laser Physics (Yuen 1976). The scheme described above can be illustrated with explicit evaluation of the values attached to the nodes for specific values of $\bf p$. 

For example if {\bf p=1} for which $\lambda_j = (j+1)$ and
\begin{equation}
{\tilde G_n}(1;y)\ =\ G_n(1,1;1;y)\ =\ \sech\ y\  \ (\tanh y)^n\ \ ,\ \ n=0,1,2,.....\ .  \label{}
\end{equation}
The expansion of these functions in powers of $y$ leads to integer coefficients after factorial factors are extracted and the set of integer coefficients may be captured in a scheme of the form
$$
\begin{array}{ccccccccccc}
{\bf 1}&&&&&&{  }\\
& \stackrel{1}{\searrow}&&&&&{  }\\
&&{\bf 1}&&&&{  }\\
& \stackrel{1}{\swarrow}&&\stackrel{2}{\searrow}&&&{ }\\
{\bf 1} &&&& {\bf 2} &&{ }\\
& \stackrel{1}{\searrow}&&\stackrel{2}{\swarrow}&&\stackrel{3}{\searrow}&{ }\\
&&{\bf 5}&&&&{\bf 6}\\
& \stackrel{1}{\swarrow}&&\stackrel{2}{\searrow}&&\stackrel{3}{\swarrow}&{ }\\
{\bf 5}&&&&{\bf 28}&&{ }\\
& \stackrel{1}{\searrow}&&\stackrel{2}{\swarrow}&&&{ }\\
&&{\bf 61}&&&&{ }\\
&\stackrel{1}{\swarrow}&&&&&{ }\\
{\bf 61}&&&&&&{ }\\
\end{array}
$$
which is reminiscent of the Pascal triangle for the binomial coefficients. The weight factors associated with the links are simply the sequence of $\lambda_j$s when viewed along the diagonals either going down the right or down the left. In this diagram the numbers at the nodes of the different columns relate to the power series representations of ${\tilde G_n}$ as given by
\begin{align}
{\tilde G_0}(1;y)\ &=\ \sech\ y\ =\ 1\ -\ \frac{y^2}{2!}\ +\ 5\ \frac{y^4}{4!}\ -\ 61\ \frac{y^6}{6!}\ +\ .. \\ 
{\tilde G_1}(1;y)\ &=\ \sech\ y\ \tanh y\ =\ y\ -\ 5\ \frac{y^3}{3!}\ +\ 61\ \frac{y^5}{5!}\ -\ ..  \\
{\tilde G_2}(1;y)\ &=\ \sech\ y\ \tanh^2 y\ =\ 2\frac{y^2}{2!}\ -\ 28\ \frac{y^4}{4!}\ +\ ..  \label{}
\end{align}
In general the $n^{th}$ column of the triangle scheme can be related to the power series for ${\tilde G_{n-1}}(1;y)$. It is simple to verify that
\begin{equation}
\frac{\partial}{\partial y}\ {\tilde G_{n+1}}(1;y)\ =\ (n+1)\ {\tilde G_{n}}(1;y)\ -\ (n+2)\ {\tilde G_{n+2}}(1;y) \ .\label{}
\end{equation}
The numbers at the nodes in the first column of the diagram are the Euler numbers. To find higher order terms in the power series expansions the diagram can be extended by proceeding further along the outer diagonal thereby generating a larger diagram.

\medskip

As the next simple we example consider the case when {\bf p=2} for which $\lambda^2_{j} = (j+1)(j+2)$. Now the resulting functions
\begin{equation}
G_n(1,2;1;y)\ =\ G_n(2,1;1;y)\ =\ {\sqrt{n+1}}\ \sech^2\ y\ (\tanh y)^n\ \ ,\ \ n=0,1,2,.....\   \label{}
\end{equation}
satisfy the recursion relation
\begin{equation}
\frac{\partial}{\partial y}G_{n+1}\ =\ {\sqrt{n+2}}\ \Big({\sqrt{n+1}}\ G_n\ -\ {\sqrt{n+3}}\ G_{n+2}\Big)\ .\label{}
\end{equation}
The factor ${\sqrt{n+1}}$ present in these functions may be removed if we define
\begin{align}
{\tilde G}_n(2;y)\ &=\ \frac{1}{\sqrt{n+1}}\ G_n(1,2;y)\  =\ \sech^2\ y\ \tanh^n y\\
{\bar G}_n(2;y)\ &=\ \ {\sqrt{n+1}}\ G_n(1,2;y)\  =\ (n+1)\ \sech^2\ y\ \tanh^n y \label{}
\end{align}
which satisfy the recursion relations
\begin{align}
\frac{\partial }{\partial y}\ {{\tilde G}_{n+1}(2:y)}\  &=\ (n+1)\ {\tilde G}_{n}(2:y)\ -\ (n+3)\ {\tilde G}_{n+2}(2:y)  \\
\frac{\partial }{\partial y}\ {{\bar G}_{n+1}(2:y)}\  &=\ (n+2)\ \Big({\bar G}_{n}(2:y)\ -\ {\bar G}_{n+2}(2:y)\Big) \ .\label{}
\end{align} 
The expansion of ${\tilde G}_n$ and ${\bar G}_n$ in powers of $y$ lead to integer coefficients after factorial factors are extracted and the resulting set of integer coefficients may be characterized by triangular schemes of the form
$$
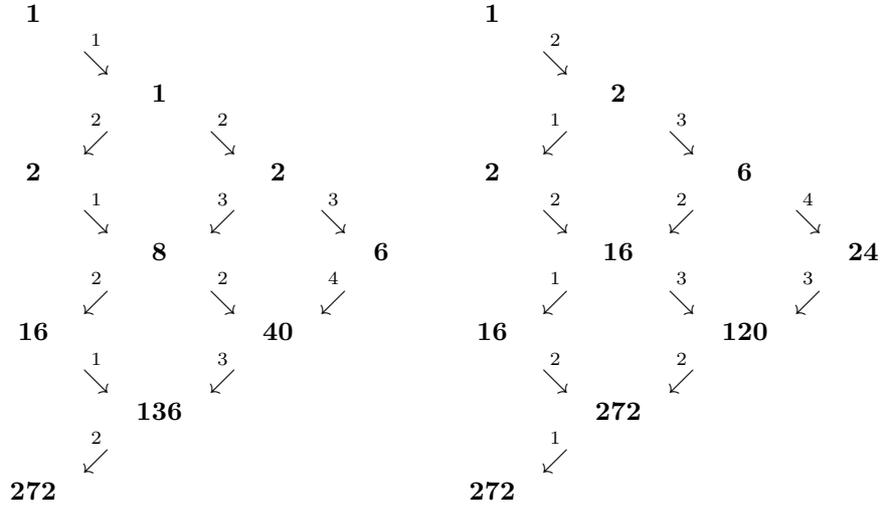
\begin{array}{ccccccccccc}
{\bf 1}&&&&&&{  }\\
& \stackrel{1}{\searrow}&&&&&{  }\\
&&{\bf 1}&&&&{  }\\
& \stackrel{2}{\swarrow}&&\stackrel{2}{\searrow}&&&{ }\\
{\bf 2} &&&& {\bf 2} &&{ }\\
& \stackrel{1}{\searrow}&&\stackrel{3}{\swarrow}&&\stackrel{3}{\searrow}&{ }\\
&&{\bf 8}&&&&{\bf 6}\\
& \stackrel{2}{\swarrow}&&\stackrel{2}{\searrow}&&\stackrel{4}{\swarrow}&{ }\\
{\bf 16}&&&&{\bf 40}&&{ }\\
& \stackrel{1}{\searrow}&&\stackrel{3}{\swarrow}&&&{ }\\
&&{\bf 136}&&&&{ }\\
&\stackrel{2}{\swarrow}&&&&&{ }\\
{\bf 272}&&&&&&{ }\\
\end{array}
 \quad\quad
\begin{array}{ccccccccccc}
{\bf 1}&&&&&&{  }\\
& \stackrel{2}{\searrow}&&&&&{  }\\
&&{\bf 2}&&&&{  }\\
& \stackrel{1}{\swarrow}&&\stackrel{3}{\searrow}&&&{ }\\
{\bf 2} &&&& {\bf 6} &&{ }\\
& \stackrel{2}{\searrow}&&\stackrel{2}{\swarrow}&&\stackrel{4}{\searrow}&{ }\\
&&{\bf 16}&&&&{\bf 24}\\
& \stackrel{1}{\swarrow}&&\stackrel{3}{\searrow}&&\stackrel{3}{\swarrow}&{ }\\
{\bf 16}&&&&{\bf 120}&&{ }\\
& \stackrel{2}{\searrow}&&\stackrel{2}{\swarrow}&&&{ }\\
&&{\bf 272}&&&&{ }\\
&\stackrel{1}{\swarrow}&&&&&{ }\\
{\bf 272}&&&&&&{ }\\
\end{array}
$$ 
The numbers at the nodes of the different columns of the diagram on the left relate to the power series representations of ${\tilde G}_n(2:y)$ as given by
\begin{align}
{\tilde G_0}(2:y)\ &=\ \sech^2\ y\ =\ 1\ -\ 2\ \frac{y^2}{2!}\ +\ 16\ \frac{y^4}{4!}\ -\ 272\ \frac{y^6}{6!}\ +\ .. \\ 
{\tilde G_1}(2:y)\ &=\ \sech^2\ y\ \tanh y\ =\ y\ -\ 8\ \frac{y^3}{3!}\ +\ 136\ \frac{y^5}{5!}\ -\ ..\\
{\tilde G_2}(2:y)\ &=\ \sech^2\ y\ \tanh^2 y\ =\ 2\ \frac{y^2}{2!}\ -\ 40\ \frac{y^4}{4!}\ ..  \label{}
\end{align}
In general the $n^{th}$ column of the triangle scheme can be related to the power series for ${\tilde G}_{n-1}(y)$. The numbers at the nodes in the first column of this diagram are the Tangent numbers which are simply related to the Bernoulli numbers.

\medskip

When $\sigma \ne 1$, but a finite number, it is simply a scale factor and the resulting functions may be obtained using a scaled variable instead of $y$. However, when $\sigma=0$ but $\lambda_j \ne 0$, the functions that arise in this limit have a new structure and must be studied separately. This is the case we examine next.

\subsection{Study of the case $\sigma = 0$}

We consider the case when one of $(\alpha,\beta)$ takes the value 1, the other $\rightarrow \infty$, $\sigma \rightarrow 0$ and $\sigma \alpha \beta\ \rightarrow 1$. In this limit $\lambda_j^2\ \rightarrow (1 +j)$, and the commutator $[L,R] \rightarrow I$, the unit matrix. Rowe  {\it et al.} (2001) has also studied how the group functions behave in the limit $\sigma \rightarrow 0$ and has shown that the limit process is not sensitive to the use of su(1,1) or su(2) functions as a starting point ({\it i.e}) whether the limit $\sigma\rightarrow 0$ is approached from positive or negative values of $\sigma$. 
The generators corresponding to this algebra are the raising and lowering operators of the SHO ({\it i.e}), $L = a\ ,\ R = a^{\dagger}$. Using the relation 
\begin{equation}
Lt_{\sigma\rightarrow 0}\ \big({\sech\ y{\sqrt{\sigma}}}\big)^{\frac{1}{\sigma}}\ \rightarrow\ \exp({\frac{-y^2}{2}})  \label{}
\end{equation}
in (\ref{eq:n1} - \ref{eq:n3}) leads to the result that
\begin{equation}
{\exp(iy(a\ +\ a^{\dagger}))}\ =\ \big({\exp(iya^{\dagger})}\big)\  \big({\exp( \frac{-y^2}{2}})\big)\ \big({\exp(iya)}\big) \label{}
\end{equation}
which is well known in laser studies as the Unitary operator which may be used to generate the coherent states of the electromagnetic field  from the vacuum state. It may be established that
\begin{align}
G_n\  &\rightarrow \ \frac{y^n}{\sqrt{n!}}\ \exp {\frac{-y^2}{2}} \\
\frac{\partial}{\partial y}G_{n+1}\ &=\ \big({\sqrt{n+1}}\ G_n\ -\ {\sqrt{n+2}}\ G_{n+2}\big)\ .\label{}
\end{align}
The diagram corresponding to $G_n$ has square root factors in the arms and nodes. Simpler diagrams result if we consider
\begin{equation}
{\tilde G}_n\ =\ \frac{G_n}{\sqrt{{n!}}}\ =\ \frac{y^n}{n!}\ \exp({\frac{-y^2}{2}}) \quad\quad {\hbox {or}}\quad\quad
{\bar G}_n\ =\ {\sqrt{n!}}\ G_n\ =\ y^n\ \exp({\frac{-y^2}{2}}) \label{}
\end{equation}
whose power series expansion is coded by the nodes of 
$$
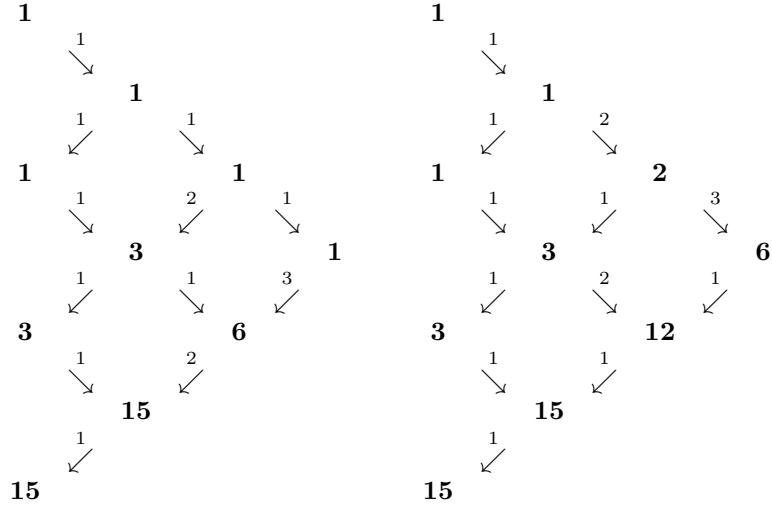
\begin{array}{ccccccccccc}
{\bf 1}&&&&&&{  }\\
& \stackrel{1}{\searrow}&&&&&{  }\\
&&{\bf 1}&&&&{  }\\
& \stackrel{1}{\swarrow}&&\stackrel{1}{\searrow}&&&{ }\\
{\bf 1} &&&& {\bf 1} &&{ }\\
& \stackrel{1}{\searrow}&&\stackrel{2}{\swarrow}&&\stackrel{1}{\searrow}&{ }\\
&&{\bf 3}&&&&{\bf 1}\\
& \stackrel{1}{\swarrow}&&\stackrel{1}{\searrow}&&\stackrel{3}{\swarrow}&{ }\\
{\bf 3}&&&&{\bf 6}&&{ }\\
& \stackrel{1}{\searrow}&&\stackrel{2}{\swarrow}&&&{ }\\
&&{\bf 15}&&&&{ }\\
&\stackrel{1}{\swarrow}&&&&&{ }\\
{\bf 15}&&&&&&{ }\\
\end{array}
 \quad\quad
\begin{array}{ccccccccccc}
{\bf 1}&&&&&&{  }\\
& \stackrel{1}{\searrow}&&&&&{  }\\
&&{\bf 1}&&&&{  }\\
& \stackrel{1}{\swarrow}&&\stackrel{2}{\searrow}&&&{ }\\
{\bf 1} &&&& {\bf 2} &&{ }\\
& \stackrel{1}{\searrow}&&\stackrel{1}{\swarrow}&&\stackrel{3}{\searrow}&{ }\\
&&{\bf 3}&&&&{\bf 6}\\
& \stackrel{1}{\swarrow}&&\stackrel{2}{\searrow}&&\stackrel{1}{\swarrow}&{ }\\
{\bf 3}&&&&{\bf 12}&&{ }\\
& \stackrel{1}{\searrow}&&\stackrel{1}{\swarrow}&&&{ }\\
&&{\bf 15}&&&&{ }\\
&\stackrel{1}{\swarrow}&&&&&{ }\\
{\bf 15}&&&&&&{ }\\
\end{array}
$$ 
where the columns on the left diagram code for the power series of ${\tilde G}_n$ and the columns on the right diagram code for ${\bar G}_n$. For example,
\begin{align}
G_0\ =\ {\tilde G}_0\ &=\ {\bar G}_0\ =\ 1\ - \frac{y^2}{2!}\ +\ 3\ \frac{y^4}{4!}\ -\ 15\ \frac{y^6}{6!}\ +\ ....    \\
G_1\ =\ {\tilde G}_1\ &=\ {\bar G}_1\ =\ y\ - 3\ \frac{y^3}{3!}\ +\ 15\ \frac{y^5}{5!}\ .... \\
\frac{G_2}{{\sqrt{2!}}} =\ {\tilde G}_2\ &=\ \frac{{\bar G}_n}{2!}\ =\ \frac{y^2}{2!}\ -\ 6\ \frac{y^4}{4!}\ +\ .... \label{}
\end{align}  

\subsection{Generalization to diagrams for particle and hole states}

The analysis of the cases considered so far indicates that for general values of $\alpha$ and $\beta$ the functions defined by (\ref{eq:f1}) would correspond to functions generated from the nodes of a class of diagrams which have a systematic pattern of weights along the diagonals going down to the left or to the right. If neither of $(\alpha,\beta)$ equals $1$ the 'particle' and 'hole' states no longer decouple and the triangular diagrams are no longer adequate to express the functions $G_n$ that arise from the general algebra. However a triangle diagram reflected about the vertical axis produces a diamond formation and such diagrams may be used to accommodate the feature that the nodes to the right of the central vertical line may be used for describing the 'particle' states and the nodes to the left of the central spine may be used for the 'hole' states.
$$
\begin{array}{cccccccccccccc}
&&&&&&{\bullet}&&&&&&{  }\\
&&&&& \stackrel{\lambda_{-1}}{\swarrow}&& \stackrel{\lambda_0}{\searrow}&&&&&{  }\\
&&&&{\bullet}&&&&{\bullet}&&&&{  }\\
&&& \stackrel{\lambda_{-2}}{\swarrow}&&\stackrel{\lambda_{-1}}{\searrow}&&\stackrel{\lambda_0}{\swarrow}&&\stackrel{\lambda_1}{\searrow}&&&{ }\\
&& {\bullet}&&&&{\bullet} &&&& {\bullet}&&{ }\\
& \stackrel{\lambda_{-3}}{\swarrow}&&\stackrel{\lambda_{-2}}{\searrow}&&\stackrel{\lambda_{-1}}{\swarrow}&&\stackrel{\lambda_0}{\searrow}&&\stackrel{\lambda_1}{\swarrow}&&\stackrel{\lambda_2}{\searrow}&{ }\\
{\bullet}&&&&{\bullet}&&&&{\bullet}&&&&{\bullet}\\
&\stackrel{\lambda_{-3}}{\searrow}&&\stackrel{\lambda_{-2}}{\swarrow}&&\stackrel{\lambda_{-1}}{\searrow}&&\stackrel{\lambda_0}{\swarrow}&&\stackrel{\lambda_1}{\searrow}&&\stackrel{\lambda_2}{\swarrow}&{ }\\
&&{\bullet}&&&&{\bullet}&&&&{\bullet}&&{ }\\
&&& \stackrel{\lambda_{-2}}{\searrow}&&\stackrel{\lambda_{-1}}{\swarrow}&&\stackrel{\lambda_0}{\searrow}&&\stackrel{\lambda_1}{\swarrow}&&&{ }\\
&&&&{\bullet}&&&&{\bullet}&&&&{ }\\
&&&&&\stackrel{\lambda_{-1}}{\searrow}&&\stackrel{\lambda_0}{\swarrow}&&&&&{ }\\
&&&&&&{\bullet}&&&&&&{ }\\
\end{array}
$$
It is evident from this diagram that when $\lambda_{-1}=0$ the weights attached to the links in the first column to the left of the central spine all vanish and the two halves of the diagram uncouple and the values at the nodes on the central spine and to the right hand side of the spine are no longer affected by the left side of the diagram and leads to the triangular diagrams we have considered earlier. Similar decoupling of the diagram into two parts will happen whenever one of the $\lambda_j$s $\rightarrow 0$. We next consider an example of an algebra which leads to functions coded by diagrams in diamond formation.

\medskip

If the parameters of the algebra are chosen so that $\alpha = \beta\ \rightarrow\ \infty ,\ \sigma \rightarrow \ 0,\ \alpha\beta \sigma \rightarrow \ 1$, in this limit $\lambda_j\ \rightarrow\ 1$  independent of $j$,  then $S$ vanishes because $R$ and $L$ now commute and in (\ref{eq:n3}) $f=1,\ g=1$, and (\ref{eq:f2}) simplifies. The resulting solutions
\begin{equation}
G_n\ =\ \sum^{\infty}_{j=0}\ (-)^j\ \frac{y^{2j+n}}{j!(j+n)!}\ =\ J_n(2y) \label{}
\end{equation}
may be identified as the Bessel functions of integer orders (Abramowitz and Stegun 1965) that satisfy
\begin{equation}
\frac{\partial}{\partial y}J_{n+1}(2y)\ =\ \big(\ J_n(2y)\ -\ J_{n+2}(2y)\big) \label{eq:j1}
\end{equation}
which is the $\lambda_j\ \rightarrow\ 1$ limit of (\ref{eq:r1}). The power series expansion of these solutions may be represented by the diagram
$$
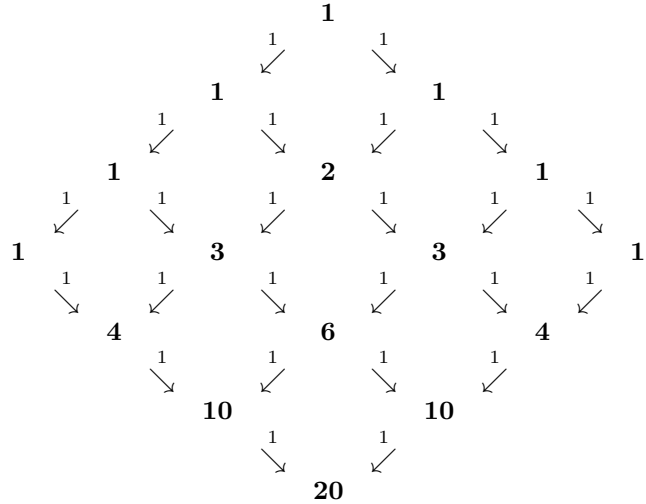
\begin{array}{cccccccccccccc}
&&&&&&{\bf 1}&&&&&&{  }\\
&&&&& \stackrel{1}{\swarrow}&& \stackrel{1}{\searrow}&&&&&{  }\\
&&&&{\bf 1}&&&&{\bf 1}&&&&{  }\\
&&& \stackrel{1}{\swarrow}&&\stackrel{1}{\searrow}&&\stackrel{1}{\swarrow}&&\stackrel{1}{\searrow}&&&{ }\\
&& {\bf 1}&&&&{\bf 2} &&&& {\bf 1}&&{ }\\
& \stackrel{1}{\swarrow}&&\stackrel{1}{\searrow}&&\stackrel{1}{\swarrow}&&\stackrel{1}{\searrow}&&\stackrel{1}{\swarrow}&&\stackrel{1}{\searrow}&{ }\\
{\bf 1}&&&&{\bf 3}&&&&{\bf 3}&&&&{\bf 1}\\
&\stackrel{1}{\searrow}&&\stackrel{1}{\swarrow}&&\stackrel{1}{\searrow}&&\stackrel{1}{\swarrow}&&\stackrel{1}{\searrow}&&\stackrel{1}{\swarrow}&{ }\\
&&{\bf 4}&&&&{\bf 6}&&&&{\bf 4}&&{ }\\
&&& \stackrel{1}{\searrow}&&\stackrel{1}{\swarrow}&&\stackrel{1}{\searrow}&&\stackrel{1}{\swarrow}&&&{ }\\
&&&&{\bf 10}&&&&{\bf 10}&&&&{ }\\
&&&&&\stackrel{1}{\searrow}&&\stackrel{1}{\swarrow}&&&&&{ }\\
&&&&&&{\bf 20}&&&&&&{ }\\
\end{array}
$$
The central spine corresponds to $n=0$ and columns to the right of the spine correspond to positive integer values of $n$ and the columns to the left of the spine correspond to negative integer values of $n$. The upper part of the diagram is the Pascal triangle with the nodes along the horizontal rows providing the binomial coefficients. We now have the additional insight that the numbers at the nodes of the columns may be used to provide :
\begin{align}
J_0(2y)\ &=\ 1\ -\ 2\ \frac{y^2}{2!}\ +\ 6\ \frac{y^4}{4!}\ -\ 20\ \frac{y^6}{6!}\ +..\ &&=\ \frac{1}{0!0!}\ -\ \frac{y^2}{1!1!}\ +\ \frac{y^4}{2!2!}\ -\ \frac{y^6}{3!3!}\ +\ ....     \\
J_1(2y)\ &=\ y\ -\ 3\ \frac{y^3}{3!}\ +\ 10\ \frac{y^5}{5!}\ -\ ..\ &&=\ \frac{y}{0!1!}\ -\ \frac{y^3}{1!2!}\ +\ \frac{y^5}{2!3!}\ -\ ...\ .   \label{}
\end{align}
In this manner the columns of the diagram, when extended, provide the elements for constructing the power series for Bessel functions, the $n^{th}$ column to the right of the spine codes for $J_n(2y)$ and the columns to the left of the spine code for $(-)^{n}J_{-n}(2y)$. Using the properties of the binomial coefficients it is evident that the values at the nodes in the $m^{th}$ row add to $2^{m-1}$ when added with the same sign and add to zero for all rows except for the first row, if added with alternating signs. These features together with a property of the Bessel functions, ({\it viz.}) $J_{-n}(z) = (-)^n J_n(z)$, can be used to infer from the diagram that
\begin{align}
J_0(2y)\ +\ 2\ \sum^{\infty}_{k=1}\ J_{2k}(2y)\ &=\ 1 \\
J_0(2y)\ +\ 2\ \sum^{\infty}_{k=1}\ (-)^k\ J_{2k}(2y)\ &=\ 1\ -\ \frac{2^2 y^2}{2!}\ +\ \frac{2^4 y^4}{4!}\ -\ ...\ &&=\ {\cos 2y} \\
2\ \sum^{\infty}_{k=1}\ (-)^{k+1}\ J_{2k-1}(2y)\ &=\ 2y\ -\ \frac{2^3 y^3}{3!}\ +\ \frac{2^5 y^5}{5!}\ -\ ...\ &&=\ {\sin 2y} \label{}
\end{align}
which are well known sum rules for Bessel functions (Gradshteyn and Ryzhik 1965). This is an example of how  diagrams may be used not only to capture the power series for a whole class of functions but also extract sum rules satisfied by the functions.

\section{Generalization of Unitary operator}

The normal ordered form of $U_1$ given in (\ref{eq:n1}) is a special case of a more general operator constructed as the exponential of $(a L\ +\ bR\ +\ cS)$ where $(a,b,c)$ are complex numbers. The normal ordered decomposition can be performed using a generalization of the method used to obtain the result given in (\ref{eq:n1}). For the algebra governed by (\ref{eq:Al}) it may be shown that
\begin{align}
U_2\ &=\ \big(\exp( a L\ +\ b R\ +\ cS)\big)\ =\ (\exp(bf_+R))\ \Big(\sum_j |j\rangle \ (g_+^{\frac{S_{jj}}{\sigma}})\ \langle j|\Big)\ (\exp(af_+L)) \label{eq:n5} \\
&=\ (\exp(af_-L))\ \Big(\sum_j (|j\rangle\ (g_-^{-\frac{S_{jj}}{\sigma}})\ \langle j|\Big)\ (\exp(bf_-R)) \label{eq:n6} \\
f_{\pm}\ &=\ \frac{\tan q}{(\ q\ \mp\ c\ \sigma\ \tan q)}\ ,\ g_{\pm}\ =\ \frac{q\ \sec q}{(\ q\ \mp\ c \sigma\ \tan q\ )}\ \ ,\ \ q\ =\ {\sqrt{a\ b\ \sigma\ -\ c^2\ \sigma^2}}\   .\label{eq:n7}
\end{align}
For the case $ \sigma \ne 0,\ a = b = iy,\ c = 0$, $U_2$ reduces to $U_1$ and (\ref{eq:n5}) - (\ref{eq:n7}) reduce to (\ref{eq:n1}) - (\ref{eq:n3}). 

The factorized form of $U_2$ may be used to obtain the  rotation matrices for the angular momentum algebra for which $(R,L,S)\ \Rightarrow \big(J_+,J_-,J_z)$ and $\sigma = -\frac{1}{2}$. Choice of the coefficients $(a,b,c)$ to be related to the Cartesian components of a vector ${\bf W}$ with polar coordinates $(\omega,\theta,\phi)$ by the relations
\begin{equation}
a\ =\ i\sqrt{2}\ \omega\ \sin{\theta}\ (\exp{-i\phi})\ ,\ b\ =\ i\sqrt{2}\ \omega\  \sin{\theta}\ (\exp{i\phi})\ ,\ c\ =\ -2 i \omega\ \cos{\theta}\label{}
\end{equation}
leads to the normal and anti-normal ordered forms
\begin{align}
U\ &=\ \big(\exp(2i {\bf{W.J}}) \big)\\
\ &=\ \big(\exp(i J_+\ h\ e^{-i\phi})\big)\ \Big(\sum^{m=j}_{m=-j} |jm\rangle\ \big( s \big)^{-2m} \langle jm| \Big)\ \big(\exp(i J_-\ h\ e^{+i\phi})\big) \\
\ &=\ \big(\exp(i J_-\ h^{\star} e^{+i\phi})\big)\ \Big(\sum^{m=j}_{m=-j} | jm\rangle \ \big( s^{\star}\big)^{+2m} \langle jm|\ \Big) \big(\exp(i J_+\ h^{\star}\ e^{-i\phi})\big) \\
h\ &=\ \frac{{\sqrt {2}}}{s}\ {\sin \theta}\ {\sin \omega} \ ,\ \quad s\ =\ \big(\cos \omega\  -\ i\ \cos \theta\ \sin \omega\big) \label{}
\end{align}
where $|jm\rangle$ are eigenstates of ${\bf J}^2$ and $J_z$ with eigenvalues $j(j+1)$ and $m$ respectively. The matrix elements of the Unitary operator $U$ may be computed for any ${\bf W}$ and ${\bf J}$. For example, if ${\bf J}=1$ then using the above representations it can be established that
\begin{equation}
U \left[\begin{array}{cc} |1,-1\rangle \\ \\|1,\ \ 0\rangle\\ \\|1,+1\rangle\end{array}\right]\ =\ \left[\begin{matrix}s^2 &i\ h\ s^2\ e^{-i\phi} &- \frac{1}{2!}\  h^2\ s^2\ e^{-2i\phi} \\ \\
i\ h\ s^2\ e^{i\phi} & 1\ -\ h^2\ s^2 &i\ h^{\star}\ {s^{\star}}^2\ e^{-i\phi}\\ \\
- \frac{1}{2!}\ {h^{\star}}^2\ {s^{\star}}^2\  e^{2i\phi} &i\ h^{\star}{s^{\star}}^2 \ e^{i\phi} &{s^{\star}}^2\end{matrix}\right]\left[\begin{array}{cc} |1,-1\rangle \\ \\|1,\ \ 0\rangle\\ \\|1,+1\rangle\end{array}\right]
\end{equation}
which provides the matrix elements of $U$ for arbitrary rotations ${\bf W}$ of the ${\bf J}=1$ system. Furthermore, if $\theta=\pi/2,\ \phi=\pi/2$, then
\begin{equation}
U\ =\ \exp (-2i\omega J_y)\  =\ \left[\begin{matrix} \cos^2\omega  &{\sqrt 2}\ {\sin \omega}\ {\cos \omega} &\sin^2\omega\\ \\
-{\sqrt 2}\ {\sin \omega}\ {\cos \omega} & 1-2\sin^2\omega &{\sqrt 2}\ {\sin \omega}\ {\cos \omega}\\ \\
\sin^2\omega  &-{\sqrt 2}\ {\sin \omega}\ {\cos \omega} &\cos^2\omega \end{matrix}\right]
\end{equation}
in agreement with a well known result (Sakurai 1994).

\section{Phase operators and Unitary evolution}

A special case of the algebra studied in this paper arises when $\lambda_{-1}=0$ and $\lambda_j=1$ for $j=0$ and all positive integral values of $j$. In this case the 'particle' and 'hole' states decouple and $S_{00}=1$ and all other matrix elements of $S$ vanish. The operators $L$ and $R$ for this case may be identified with the operators $P$ and $P^{\dagger}$ introduced in the context of the study of the phase of the electromagnetic field (Susskind and Glogower 1964, Carruthers and Nieto 1965). The phase operators considered by Carruthers and Nieto satisfy the relations
\begin{align}
P\ |0\rangle\ &=\ 0\quad,\quad P\ |n\rangle\ = |n-1\rangle\quad,\quad P^{\dagger}\  |n-1\rangle\ =\ |n\rangle\quad ,\quad n \geq 1 \\
[P\ ,\ P^{\dagger}] |0\rangle\ &=\ |0\rangle\quad\quad [P\ ,\ P^{\dagger}] |n\rangle\ =\ 0\quad,\quad n\geq 1 \ .\label{}
\end{align}
For this case the normal ordering of an operator in the form of an exponential of $(a P\ + b P^{\dagger})$ as in ($\ref{eq:n1}$) is not possible because of the fractured  structure of the commutation relation of $P$ and $P^{\dagger}$ which has a  non-vanishing vacuum expectation value but all other matrix elements vanish. However, it has been shown (Sukumar 1989) that
\begin{equation}
\langle n|\ \big(\exp iy(P + P^{\dagger})\big)\ |m\rangle\ =\ i^{n-m}\ \big( J_{n-m}(2y)\ +\ (-)^m\ J_{n+m+2}(2y)\big) \label{}
\end{equation}
where  $J_k$ are Bessel functions. In particular
\begin{equation}
G_n(y)\ =\ (-i)^n\ \langle n|\ \big(\exp iy(P + P^{\dagger})\big)\ |0\rangle\ =\ \frac{n+1}{y}\ J_{n+1}(2y) \label{}
\end{equation}
which satisfy a recursion relation of the same form as (\ref{eq:j1}):
\begin{equation}
\frac{\partial}{\partial y}G_{n+1}(y)\ =\ \big(\ G_n(y)\ -\ G_{n+2}(y)\big) \ .\label{}
\end{equation}
The power series for $G_n$ can be captured by the triangular diagram
$$
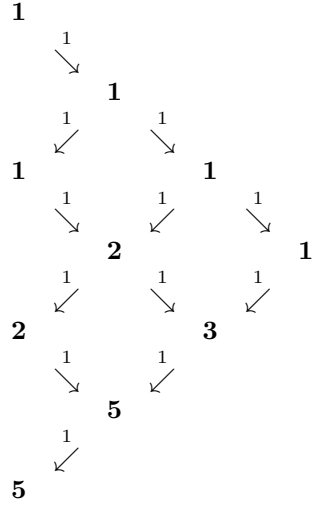
\begin{array}{ccccccccccc}
{\bf 1}&&&&&&{  }\\
& \stackrel{1}{\searrow}&&&&&{  }\\
&&{\bf 1}&&&&{  }\\
& \stackrel{1}{\swarrow}&&\stackrel{1}{\searrow}&&&{ }\\
{\bf 1} &&&& {\bf 1} &&{ }\\
& \stackrel{1}{\searrow}&&\stackrel{1}{\swarrow}&&\stackrel{1}{\searrow}&{ }\\
&&{\bf 2}&&&&{\bf 1}\\
& \stackrel{1}{\swarrow}&&\stackrel{1}{\searrow}&&\stackrel{1}{\swarrow}&{ }\\
{\bf 2}&&&&{\bf 3}&&{ }\\
& \stackrel{1}{\searrow}&&\stackrel{1}{\swarrow}&&&{ }\\
&&{\bf 5}&&&&{ }\\
&\stackrel{1}{\swarrow}&&&&&{ }\\
{\bf 5}&&&&&&{ }\\
\end{array}
$$
which gives
\begin{align}
G_0(y) &= \frac{J_1(2y)}{y} = 1\ -\ \frac{y^2}{2!}\ +\ 2\ \frac{y^4}{4!}\ -\ 5\ \frac{y^6}{6!}\ +..= \frac{1}{0!1!}\ -\ \frac{y^2}{1!2!}\ +\ \frac{y^4}{2!3!}\ -\ \frac{y^6}{3!4!}\ + .. \\
G_1(y) &= \frac{2J_2(2y)}{y} = y\ -\ 2\ \frac{y^3}{3!}\ +\ 5\ \frac{y^5}{5!}\ + ..=\ \frac{2y}{0!2!}\ -\ \frac{2y^3}{1!3!}\ +\ \frac{2y^5}{2!4!}\ - ..\ .   \label{}
\end{align}
The elements in the rows sum to easily recognizable values when added with the same sign or when added with alternating signs and these recognizable values lead to the sum rules
\begin{align}
\frac{1}{y}\ \sum^{\infty}_{k=0}\ (2k+1)\ J_{2k+1}(2y)\ &=\ 1 \label{eq:B1} \\
\frac{1}{y}\ \sum^{\infty}_{k=1}\ (2k)\ J_{2k}(2y)\ &=\ \frac{y}{1!}\ -\ \frac{y^3}{3!}\ +\ \frac{2y^5}{5!}\ -\ \frac{5y^7}{7!}\ +\  ....\  =\ \int^{y}_0\ \frac{J_1(2z)}{z}\ dz \ .\label{}
\end{align}
The first of these sum rules is listed by Gradshteyn and Ryzhik (1965) but not the second.

So far all the diagrams we have discussed correspond to the $(n0)$ matrix elements of a Unitary operator. It is also possible to draw diagrams corresponding to the power series of the functions that relate to the $(nm)$ matrix elements of $U$ for arbitrary values of $m$. For example, if we define
\begin{equation}
G_{nm}(y)\ =\ (-1)^{n-m}\ \langle n|\ \big(\exp iy(P + P^{\dagger})\big)\ |m\rangle\ =\ \big( J_{n-m}(2y)\ +\ (-)^m\ J_{n+m+2}(2y)\big) \label{}
\end{equation}
then the resulting functions for $n=0,1,2.., m=1$, have power series expansion coded by a diagram which starts in the column marked as $1$ and extends to the left as far as column marked as $0$,({\it i.e.}) the paths start in the column corresponding to $n=1$ and proceed downwards, left or right but never crossing the border at the column marked as $0$. For the particular Unitary operator considered here the coefficients at the nodes in the diagram given below measure the number of paths starting at $m=1$ and ending at $n=0,1,2,...$ .
\bigskip
$$
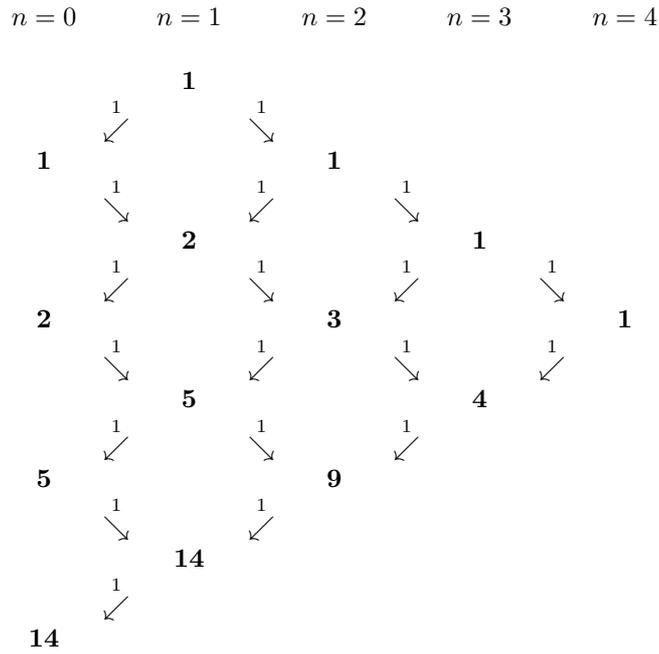
\begin{array}{ccccccccccc}
{ n=0}&&{ n=1}&&n=2&&n=3&&n=4\\
{ }&&&&&&&&\\
&&{\bf 1}&&&&&&{  }\\
& \stackrel{1}{\swarrow}&&\stackrel{1}{\searrow}&&&&&{  }\\
{\bf 1}&&&&{\bf 1}&&&&{  }\\
& \stackrel{1}{\searrow}&&\stackrel{1}{\swarrow}&&\stackrel{1}{\searrow}&&&{ }\\
&&{\bf 2} &&&& {\bf 1} &&{ }\\
& \stackrel{1}{\swarrow}&&\stackrel{1}{\searrow}&&\stackrel{1}{\swarrow}&&\stackrel{1}{\searrow}&{ }\\
{\bf 2}&&&&{\bf 3}&&&&{\bf 1}\\
&\stackrel{1}{\searrow}&& \stackrel{1}{\swarrow}&&\stackrel{1}{\searrow}&&\stackrel{1}{\swarrow}&{ }\\
&&{\bf 5}&&&&{\bf 4}&&{ }\\
&\stackrel{1}{\swarrow}&& \stackrel{1}{\searrow}&&\stackrel{1}{\swarrow}&&&{ }\\
{\bf 5}&&&&{\bf 9}&&&&{ }\\
&\stackrel{1}{\searrow}&&\stackrel{1}{\swarrow}&&&&&{ }\\
&&{\bf 14}&&&&&&{ }\\
&\stackrel{1}{\swarrow}&&&&&&&\\
{\bf 14}&&&&&&&&
\end{array}
$$

Similarly the power series expansions for the functions $G_{n2}$ are coded by the coefficients of a diagram which starts in the column marked $2$ and proceeding downwards,left or right, but never crossing the border at the column marked $0$ as shown below:
$$
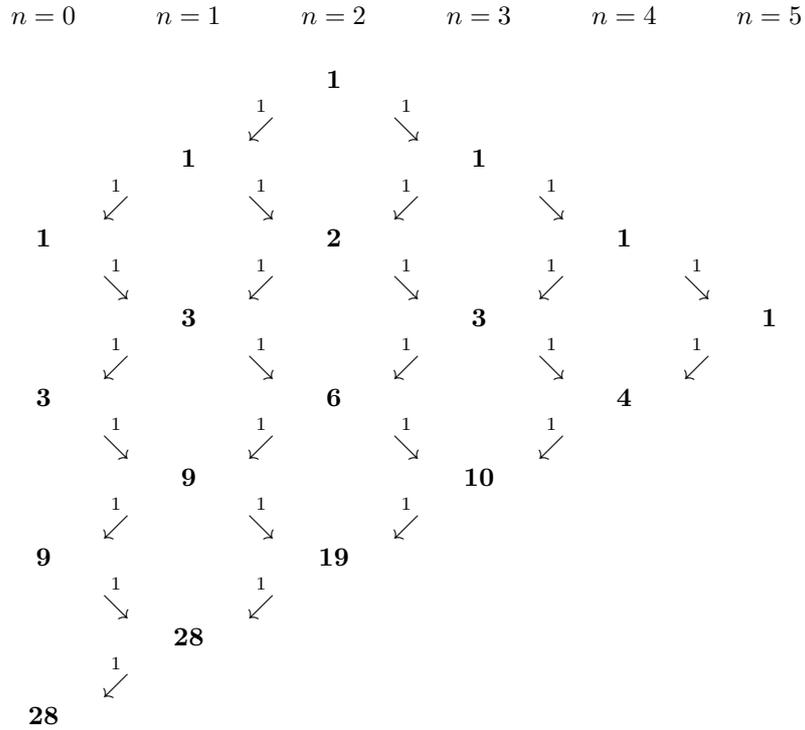
\begin{array}{ccccccccccc}
 n=0&&n=1&&n=2&&n=3&&n=4&&n=5\\
{ }&&&&&&&&&&\\
&&&&{\bf 1}&&&&&&{  }\\
&&& \stackrel{1}{\swarrow}&&\stackrel{1}{\searrow}&&&&&{  }\\
&&{\bf 1}&&&&{\bf 1}&&&&{  }\\
&\stackrel{1}{\swarrow}&&\stackrel{1}{\searrow}&&\stackrel{1}{\swarrow}&&\stackrel{1}{\searrow}&&&{ }\\
{\bf 1}&&&&{\bf 2} &&&& {\bf 1} &&{ }\\
&\stackrel{1}{\searrow}&& \stackrel{1}{\swarrow}&&\stackrel{1}{\searrow}&&\stackrel{1}{\swarrow}&&\stackrel{1}{\searrow}&{ }\\
&&{\bf 3}&&&&{\bf 3}&&&&{\bf 1}\\
&\stackrel{1}{\swarrow}&&\stackrel{1}{\searrow}&& \stackrel{1}{\swarrow}&&\stackrel{1}{\searrow}&&\stackrel{1}{\swarrow}&{ }\\
{\bf 3}&&&&{\bf 6}&&&&{\bf 4}&&{ }\\
&\stackrel{1}{\searrow}&&\stackrel{1}{\swarrow}&&\stackrel{1}{\searrow}&&\stackrel{1}{\swarrow}&&&{ }\\
&&{\bf 9}&&&&{\bf 10}&&&&{ }\\
&\stackrel{1}{\swarrow}&&\stackrel{1}{\searrow}&&\stackrel{1}{\swarrow}&&&&&{ }\\
{\bf 9}&&&&{\bf 19}&&&&&&{ }\\
&\stackrel{1}{\searrow}&&\stackrel{1}{\swarrow}&&&&&&&{ }\\
&&{\bf 28}&&&&&&&&\\
&\stackrel{1}{\swarrow}&&&&&&&&&{}\\
{\bf 28}&&&&&&&&&&{}
\end{array}
$$
Similar diagrams may be drawn to capture the coefficients that code for the power series expansion of $G_{nm}$ for any value of $m$ using the procedure outlined above by starting in the column marked with the value of $m$ and drawing all downward bound paths which do not cross the border at $n=0$.

\section{Conclusions}

In this study we have shown that a variety of solutions emerge from the expectation values of a Unitary operator constructed from the generators of an algebra which can be related to su(2) or su(1,1) depending on the value of a parameter. Coherent states and squeezed states of the SHO familiar in laser studies emerge as special cases. The rotation matrices associated with the angular momentum algebra also emerges as a special case. We have shown that the functions which arise as expectation values of the Unitary operator satisfy a simple two term recursion relation and have power series expansions which can be captured by diagrams similar to the Pascal triangle familiar in the study of binomial coefficients. 

We have studied the $(n0)$ matrix elements of the Unitary operator $U_1$ in (\ref{eq:n1}) in depth. For the general case we note that if we define
\begin{equation}
G_{nm}(\alpha,\beta;\sigma;y)\ =\ (-i)^{n-m}\ \langle n| U_1 |m \rangle  \\
\end{equation}
it may be shown that $G_{nm}$ may be evaluated by the same steps as for $G_n$ by starting from (\ref{eq:n2}) and the resulting hypergeometric function shows that $G_{nm}$ can be related to $G_n$ in (\ref{eq:f1}) - (\ref{eq:f0}) through the mapping given by
\begin{equation}
G_{nm}(\alpha,\beta;\sigma;y)\ =\ G_{(n-m)0}\big(\alpha+m,\beta+m;\sigma;y\big)\ \equiv\ G_{n-m}(\alpha+m,\beta+m;\sigma;y)\ . \label{}
\end{equation}
Diagrams that code for the power series of $G_{nm}$ may be drawn by adopting the procedure used for $G_n$ by starting in the column corresponding to $m$ and considering all paths proceeding downward with appropriate weights $\lambda_j$ attached to the arms.

\section{References}

Barut, A.O. and Fronsdal, C. 1965 On non-compact groups. II Representation of the 2+1 Lorentz group, {\it Proc.\ Roy.\ Soc.\ A} {\bf 287} 532.

Holman, W.J.\ and Biedenharn, L.C. 1966 Complex angular momenta and the groups SU(1,1) and SU(2) {\it Ann.\ Phys.\  (N.Y)} {\bf 39} 1-42.

Schwinger, J. 1965 in {\it Quantum Theory of Angular Momentum,} eds. Biedenharn, L.C. and Van Dam, H. (Academic, New York) 229.

Louisell, W.H. 1977 {\it Radiation and Noise in Quantum Electronics} (Kreiger, R.E., Huntington, N.Y).

Yurke, B., McCall, S.L.\ and Klauder, J.R.\ 1986 SU(2) and SU(1,1) Interferometers, {\it Phys. Rev.} {\bf A 33} 4033-4054.

Horvathy, P.A, Plyushchay, M.S and Valenzula, M. 2010 Bosons, fermions and anyons in the plane and supersymmetry {\it Ann. Phys.} {\bf 325} 1931-1975

Correa, F., Jakulsky, V. and Plyushchay. M.S 2009 Aharanov-Bohm effect on AdS2 and nonlinear supersymmetry of reflectionless Poschl-Teller system {\it Ann. Phys} {\bf 324} 1078-1094

Truax, D.R. 1985 Baker-Campbell-Hausdorf relation and unitarity of SU(2) and SU(1,1) squeeze operators, {\it Phys.\ Rev.\ } D {\bf 31}, 1988-1991

Ui, H. 1970 Clebsch-Gordan formula of the SU(1,1) group {\it Prog. Theor. Phys.} {\bf 44} 689-702

Sukumar, C.V. 1989 Revival Hamiltonians, phase operators and non-Gaussian squeezed states, {\it J. Mod.\ Optics} {\bf 36}, 1591-1605.

Gradshteyn, I.S. and Ryzhik, I.M 1965 {\it Table of Integrals, Series and Products} (Academic Press: New York) 1039, 974.

Yuen, H.P. 1976 Two photon coherent states of the radiation field, {\it Phys.\ Rev.\ }A {\bf 13}, 2226-2243.

Rowe, D.J., deGuise, H. and Sanders, B.C 2001 Asymptotic limits of SU(2) and SU(3) Wigner functions {\it J. Math. Phys.} {\bf 42} 2315-2342

Abramowitz, M. and Stegun,I.A. 1965 {\it Handbook of Mathematical Functions} (Dover: New York) 358.

Sakurai, J.J 1994 {\it Modern Quantum Mechanics} (Addison-Wesley: New York) 195.

Susskind, L. and Glogower, J. 1964 {\it Physics (N.Y.)} {\bf 1} 49.

Carruthers, P. and Nieto, M.M. 1965 Coherent states and the number-phase uncertainty relation, {\it Phys.\ Rev.\ Lett.} {\bf 14} 387-389.

Sukumar, C.V. 1989 Phase operators of the harmonic oscillator and suppression of a number state, {\it Phys.\ Rev.\ }A {\bf 40} 5426-9.

\end{document}